# Identifying translational science through embeddings of controlled vocabularies

Qing Ke[1]

[1] Center for Complex Network Research, Northeastern University, Boston, MA 02115, USA

q.ke@northeastern.edu

**ABSTRACT**

**Objective**

Translational science aims at "translating" basic scientific discoveries into clinical applications. The identification of translational science has practicality such as evaluating the effectiveness of investments made into large programs like the Clinical and Translational Science Awards. Despite several proposed methods that group publications—the primary unit of research output—into some categories, we still lack a quantitative way to place papers onto the full, continuous spectrum from basic research to clinical medicine.

**Methods**

Here we learn vector-representations of controlled vocabularies assigned to MEDLINE papers to obtain a Translational Axis (TA) that points from basic science to clinical medicine. The projected position of a term on the TA, expressed by a continuous quantity, indicates the term's

"appliedness." The position of a paper, determined by the average location over its terms, quantifies the degree of its "appliedness," which we term as "level score."

**Results**

We validate our method by comparing with previous techniques, showing excellent agreement yet uncovering significant variations of scores of papers in previously defined categories. The measure allows us to characterize the standing of journals, disciplines, and the entire biomedical literature along the basic-applied spectrum. Analysis on large-scale citation network reveals two main findings. First, direct citations mainly occurred between papers with similar scores. Second, shortest paths are more likely ended up with a paper closer to the basic end of the spectrum, regardless of where the starting paper is on the spectrum.

**Conclusions**

The proposed method provides a quantitative way to identify translational science.

**INTRODUCTION**

Translational science, research that translates basic scientific discoveries ("bench") into clinical applications ("bedside"), has received much emphasis recently, as exemplified by the Clinical and Translational Science Awards (CTSA) program launched by the National Institutes of Health (NIH). A consensual definition and consequent identification of translational science are helpful in evaluating the effectiveness of these programs and understanding translational pathways of drug development [1-6].

However, such a consensus has yet to be reached and existing proposals have limitations in several ways. For example, the four-phase (T1-T4) continuum of translational research in genomics [7], which moves from the T1 phase that seeks a health application from a genome-based discovery to T4 that evaluates health impact of the application in practice, is constrained within translational science and fails to provide ways to operationalize the four phases to research output. In a seminal study back in 1976, Narin et al. presented four levels of research activities, namely Basic Research, Clinical Investigation, Clinical Mix, and Clinical Observation, and assigned each biomedical journal to one of the four levels [8]. The method, though later used for research evaluation [9-10] and mapping exercises [11], did not operate at the paper level—our main focus here—and was largely based on the authors' judgment, lacking proper justification. Later methods classified papers into Narin's four categories based on title words [12-13]. Recently, Weber introduced a novel cell-animal-human triangle and placed papers onto one of the seven positions (three corners, three midpoints, and the center) in the triangle [14]. The placement is based on expert-assigned keywords called Medical Subject Headings (MeSH)

[15], the controlled vocabulary thesaurus maintained by the U.S. National Library of Medicine (NLM). He then considered papers with only cell and non-human animal (hereafter animal) related MeSH terms to be the most basic, papers with only human related terms the most applied, and papers with both types of terms in between. A similar idea was applied to NIH grants in a more recent work by Li et al. [16], who made the basic-applied dichotomy based on several dimensions such as model organisms studied and whether a grant is disease-oriented, as determined by MeSH terms associated to its abstract. Supervised learning-based approaches have also been employed to classify papers, based on conventional bag-of-words representation [17] or modern dense vector representation [18] of title and abstract. The utilization of these methods requires labor-intensive manual labeling of papers. All these methods suffer from one major limitation: They treated all papers (or grants) in the same defined category to have the same extent of "basicness," but there is a large variation of such extent, as we shall show and elaborate below.

By contrast, here we present a quantitative method to measure the degree of "basicness" of a paper. The method results in a continuous measure, which we call "level score" (LS), ranging from -1 to 1, with the value closer to -1 meaning that the paper is more orientated towards basic research and 1 more applied. Our proposal is inspired from two lines of literature. The first is Weber and Li's studies that leverage the fact that certain MeSH keywords indicate whether the research is done at the cell, animal, or human level as a hint to determine whether it is basic [14, 16]. The mere appearance of these terms, however, is not able to fully distinguish how basic of the research. For example, papers with human related terms can be about nursing research on patient care, or clinical studies about the test of the safety and effectiveness of a new drug product on human. The former case is often considered as less basic than the latter one [2, 4, 7], which fails to be captured by previous methods. This shortcoming is remedied by the second line of literature on learning vector-representations of entities (e.g., words), which allows one to explore relationships between them via simple arithmetic vector operations [19-20]. We apply these learning methods on MeSH terms to obtain, using the example mentioned above, similarities between human related terms and the rest, which in total determines the "basicness" of papers.

Extensive validations show that our results are consistent with previous methods. We apply our measure to more than 15 million papers published between 1980 and 2013 and find that LS is able to characterize to what extent papers associated with a journal, field, or the entire biomedical literature are oriented towards basic research (or clinical medicine).

## METHODS

### Data

We use a snapshot of MEDLINE, a large-scale bibliographic dataset for the biomedical research literature. Maintained by NLM, it is the primary database behind the widely used PubMed search engine. The data is publicly available [21] and contains over 25 million papers.

One prominent feature of MEDLINE is that each and every paper indexed there is associated with expert-assigned keywords called Medical Subject Headings (MeSH; Fig. 1A), controlled medical vocabularies organized into a hierarchical tree with 16 branches (A-N, V, and Z). As MEDLINE also indexes non-biomedical papers, we only consider the following 8 branches: A, B, C, D, E, G, M, and N, which are the most relevant to biomedical research, the most used, and contain the majority of MeSH terms (Table S1).

We assign journals in MEDLINE to different fields based on the NLM Catalog data [23], which contains 120 Broad Subject Terms such as Biochemistry, General Surgery, Nursing, etc.

Finally, as citation data in MEDLINE are available only for PubMed Central papers, we turn to the Web of Science (WoS) database for such data. To match a MEDLINE paper in WoS, we used a lookup table that maps PubMed ID (PMID) to Accession Number, the primary identifier used in WoS.

**Placing papers onto the basic-applied spectrum**

Our method relies on MeSH terms associated to papers and is of bottom-up: To quantify the "basicness" of a paper, we first place MeSH terms onto the basic-applied spectrum. Note that they are the most atomic piece of information available, and there are no lower-level "sub-MeSH" terms that allow us to accomplish the quantification task. We therefore use *a prior* coding, in a minimal way, in that the coding only applies to a subset of all terms. Leveraging the coding schema introduced by Weber [14], we consider basic terms as those related to the topics of (1) cell and molecular and (2) animal, treating them as equally basic; and applied terms those related to the topic of human. Operationally, cell and molecular terms are located in the subtrees rooted at the nodes "Cells," "Archaea," "Bacteria," "Viruses," "Molecular

Structure," and "Chemical Processes;" animal terms in the subtree rooted at the node "Eukaryota" (except "Humans"); human terms in the subtrees rooted at the nodes "Humans" and "Persons" (Table S2). The rational of the assignment is the commonly-adopted definition of clinical research—research that involve human subjects [24].

We then assign each term a score, also named as "level score," based on how similar it is to the basic and applied terms. This is achieved by two steps. First, we employ representation learning methods to obtain vector-representations (embeddings) of terms, so that their pairwise similarities can be easily calculated using the cosine similarity—a commonly used measure for such task. In doing so, we compute time-evolving co-occurrence matrices $M_t$ between MeSH terms (Fig. 1B), as biomedical knowledge may have been evolving over time. Specifically, for each year $t \in [1980, 2013]$, the entry $m_{ij}^t \in M_t$ represents the number of papers that were published in the 5-year time window from $t-4$ to $t$ whose associated MeSH terms contain both term $i$ and $j$. Co-occurrences capture how terms are used together among different types of papers. For example, "Freeze Fracturing" and "Cytochalasin B" are frequently used in papers about cell-level experiments, "Patient Participation" and "Professional-Patient Relations" co-occur in papers about patient care, but "Cytochalasin B" and "Patient Participation" have not been co-used (Fig. 1B).

Next, we embed each matrix $M_t$ into $d$-dimensional vector-space using LINE [17], a popular network embedding technique that seeks to minimize the Kullback-Leibler divergence of the connection probability in the vector-space given the empirical one. Here we set the embedding dimension $d$ to 10. After embedding, each term $i$ at time $t$ is associated with a real-valued vector $v_i^t \in \mathbb{R}^d$ (Fig. 1C). Note that we tested the robustness of our results by using another embedding method called GloVe [19], which has been widely used in natural language processing to learn vector-representations of words based on their co-occurrences and also used to study language bias [25]. All our results hold when using GloVe (Figs. S11-S18).

In the second step, we find in the vector-space an imaginary vector called Translational Axis (TA) that points from basic terms to applied terms (Fig. 1C). TA is represented by the vector $\bar{v}(\text{applied}) - \bar{v}(\text{basic})$, where $\bar{v}(\text{applied})$ and $\bar{v}(\text{basic})$ are the centroid of applied and basic terms, respectively, obtained by averaging their constituent vectors. We then project each term onto TA, and its LS is the cosine similarity between its vector and the TA vector (Fig. 1D). The score ranges from -1 to 1, and by construction, terms with larger scores are more oriented towards the applied end of the basic-applied spectrum. Tables S3 and S4 provide the top 10 most basic and applied terms in 1980.

Once obtaining LS of terms, the score of a paper published at year $t$ is then the average of the scores of its MeSH terms at $t$ (Fig. 1E). Out of all the 17,362,010 papers published between 1980 and 2013, we only consider the 15,693,562 (90.4%) papers whose majority of MeSH terms are included in our analysis.

**RESULTS**

**Validation**

As simple validations of the embedding and LS of MeSH terms, we find that for the two (cell/animal and human) categories of terms, within-category pairs of terms have significantly higher cosine similarity than between-category ones (Fig. S1) and that the distributions of LS of terms in the two categories are well-separated (Fig. S2), which is consistent across years (Fig. S3).

Before proceeding to validating LS of papers, we stress that all previous methods classified papers into some predefined categories, whereas ours assigns continuous value to papers. This difference leads to the validation exercise to be comparisons of distributions of LS of papers in those categories. First, we test the effectiveness of our method on a particular type of papers— papers flagged as clinical trials. These papers reported results from studies that evaluate the effectiveness of interventions on humans, and therefore are expected to be from the applied side. Fig. S4A confirms this, showing that the vast majority of clinical trial publications have LS greater than zero, with a median of 0.42 (*cf.* Fig. 2A). Note that the information of whether papers are about clinical trials was not used during the quantification process. Figs. S4B-E further display respectively the distributions of LS of Phase I, II, III, and IV clinical trial studies, indicating that the four phases of studies are progressively more oriented towards applied research, with their medians getting significantly larger (permutation test, $p < 10^{-5}$; Table S5).

Second, Weber classified papers into seven categories based on whether their MeSH terms contain cell, animal, and human related ones, and considered papers with only cell and animal terms to be most basic, papers with only human terms most applied, and papers with both types of terms in between [14]. Our results agree with Weber's analysis: We arrange these categories of papers based on their LS in the same order as Weber did (Table 1 and Fig. S5).

**Table 1.** Comparison between Weber's results [14] and ours. The first column lists the 7 classes of papers, categorized based on whether their MeSH terms contain the ones related to cell and molecular (C), animal (A), and human (H) (Table S2). We arrange the categories in the order from most basic to most applied, according to the values of research level showed in the second column that are taken from Weber's analysis (Table **1a** in ref. [14]). The research level of a category was defined as the weighted average of research level of the four prototype journals selected by Narin, and the weight was the inverse of the number of papers in each journal [14]. In the third column, we show our results of the same 7 categories, based on the median level score presented in the fourth column. Fig. S5 shows the distributions of level score of papers in each category. The last two columns report the number and percentage of papers in each category. The remaining 577,754 (3.68%) papers belong to none of the 7 categories.

| Category | Research level | Category | Median level score | # papers | % papers |
|---|---|---|---|---|---|
| C | 3.78 | CA | -0.19 | 729,412 | 4.65 |
| CA | 3.68 | C | -0.15 | 1,914,634 | 12.20 |
| CAH | 3.40 | CAH | -0.10 | 826,219 | 5.26 |
| A | 3.15 | A | -0.06 | 1,495,234 | 9.53 |
| CH | 2.85 | CH | 0.10 | 1,674,517 | 10.67 |
| AH | 2.10 | AH | 0.14 | 594,467 | 3.79 |

| H | 1.59 | H | 0.48 | 7,881,325 | 50.22 |

## Research level of journals and biomedical fields

We examine LS of papers in different journals and fields to characterize their standings along the basic-applied spectrum. To explore this, Fig. 2B shows distributions of LS of papers in selected journals. Our results are in full agreement with the four-level (L4-L1) categorization proposed by Narin et al. [8]: the four selected prototype journals—*Journal of Biological Chemistry* (*JBC*; L4), *Journal of Clinical Investigation* (*JCI*; L3), *New England Journal of Medicine* (*NEJM*; L2), and *Journal of the American Medical Association* (*JAMA*; L1) —are placed onto the basic-applied spectrum in the order from L4 to L1 by our method (Fig. 2B), and the median LS of papers published in these journals are -0.22, -0.11, 0.41, and 0.52, respectively. Fig. 2B also suggests that there are many "additional levels" between the four journals. Papers published in *Cell* have a similar distribution of LS with that of *JBC*. Multi-disciplinary journals such as *Nature* and *Science*, though publish papers from the applied side, cover more papers from the basic side, making the mode score comparable to that of *JBC* and *Cell*. Between *JCI* and *NEJM* are *Nature Medicine*, a preclinical medicine journal, *Neuropsychopharmacology*, a journal covering topics on both basic and clinical research about the brain and behavior, and *Nature Review Drug Discovery*, a journal on drug discovery and development. These results are consistent with a previous study that conceptually placed journals along the pipeline of translational medicine [26].

Fig. 2C focuses on the "basicness" of biomedical research fields, as defined by NLM as Broad Subject Terms, based on the designation of journals to fields. Cell Biology, Biochemistry, and Molecular Biology are close to the basic end of the basic-applied spectrum, whereas Nursing and Health Services Research the applied end. Between the two ends are, in the order from basic to applied, Allergy and Immunology, Physiology, Psychopharmacology, Neoplasms, General Surgery, among others. What is also noticeable from Fig. 2C is bimodal distributions of LS of papers from disciplines like Brain, Endocrinology, Psychopharmacology, Neurology, and Neoplasms. This meets with the intuition that research in these fields may have two goals at the same time—scientific discoveries and therapeutics. Neoplasms, for example, can be about both fundamental research on the identification of tumor suppressor genes and clinical research on the development of cancer drugs.

Finally, the entire biomedical research literature, as recorded in the MEDLINE dataset, exhibits a bimodal distribution of the score (Fig. 2A). This is interesting from the following aspects. First, it suggests that there exists a robust threshold ($th = 0.16$) to separate the two modes, therefore providing a natural classification between basic and applied papers. This may be appealing in many lines of enquiry, such as examining types of research conducted [27], where the basic-applied dichotomy is often used. Second, the fact that the median (0.27) is greater than the threshold $th$ indicates that biomedical research in its entirety is more towards the applied side, with the score greater than $th$ for 57.3% of papers. Third, the bimodality indicates that there may lack amounts of published translational science—research that is essential for "translating" basic scientific discoveries into clinical medicine.

In summary, while our results are consistent with previous qualitative studies, they uncover significant variations of the score of papers in previously defined categories, therefore highlighting the necessity of quantifying research level at the individual paper level. The proposed measure also allows us to characterize standing of journals, disciplines, and the entire biomedical literature along the basic-applied spectrum.

**Changes over time**

In the previous section, we have pooled all papers together to examine basicness of different categories. As biomedical knowledge evolves, research level of MeSH terms may change accordingly. Our sliding window approach naturally captures this, thus allowing us to detect temporal changes of research level. Fig. 3 illustrates some examples. We observe a steady decrease of LS for the term "Cloning, Organism" (Fig. 3A), indicating that the research has been moving in the direction of basic science. This may be due to the trend that experiments about organism cloning were more on human originally and then more on animal. LS of papers with this descriptor, correspondingly, has been decreasing (Fig. 3B). On the other hand, research about "Hepatitis A Vaccines" has been evolving towards human research. The term "Adipose Tissue, Brown" and its associated papers have been moving to the human direction since 1990s when it was found in adult humans [14].

**Citation linkage between basic and applied research**

Next, we study how papers with different LS are connected in the citation network. Are basic papers more likely to be cited by basic or applied ones? Where do citations originated from applied papers go? In this section, we only focus on the 10,118,672 MEDLINE papers included in our analysis that are also present in WoS, and there are 200,359,263 citation pairs between those papers. Fig. 4A plots, in the form of heat map, the LS of citing paper and that of cited paper for all 200 million citation pairs, where the color encodes the number of pairs. We observe two regions of very high density, corresponding to the case where citing and cited papers have similar scores. This means that basic research is more likely to cite other basic research, while applied cite other applied—a "homophilous" pairing of citing and cited papers with respect to their research levels. This observation cannot be explained by the bimodal distribution of LS of papers (Section S1; Fig. S7A).

We further quantify the pattern of homophilous pairing at the paper level. We calculate, for each paper, the mean difference $\mu$ between its LS and the scores of its references, capturing the average direction from which it absorbs previous knowledge. Fig. 4B shows that the mean difference is around zero for most papers regardless of their LS, which cannot be explained by the bimodality of LS (Fig. S7B). Another noticeable observation from Fig. 4B is an asymmetric dispersion of $\mu$ around zero for papers at both ends of spectrum but symmetric for papers in the middle. This indicates that there are basic papers that cite much more applied ones, applied papers that cite much more basic ones, and papers in between that cite papers from both directions.

Our results so far suggest that direct citations rarely occurred between basic and applied research. This raises the questions of whether they operate in separated spheres and how basic knowledge can subsequently be used in applied research. To answer them, we go beyond direct citations and characterize citation connectivity between papers that are steps away. To illustrate our method, let us for now focus on a single paper $i$ published in year $t_i$. We calculate the distances from $i$ to all other reachable papers, $T_i$, in the entire citation network. The potentially reachable papers are those published until $t_i$, denoted as $A_{t_i}$, since a paper can only cite previously-published ones. For visualization purpose, we discretize LS and let $I$ ($J$) be the binned LS value of paper $i$ ($j$) and $D$ is the binning function. We define the reachability of node $i$ to papers with value $J$ as the fraction of target papers that can be reached from $i$:

$$R_J^i = \frac{|j \in T_i : D(j) = J|}{|j \in A_{t_i} : D(j) = J|}.$$

This measure may provide an answer to a question related to the life-cycle of translational science: how many basic scientific discoveries have resulted in marketed drugs [2]. Our answer is not an exact one, which would require case-by-case studies, but it provides an upper bound reflected in the citation network. We also introduce two further measures: $L_J^i$, the average distance to reachable papers:

$$L_J^i = \frac{\sum_{j \in T_i : D(j) = J} l_{ij}}{|j \in T_i : D(j) = J|},$$

where $l_{ij}$ is the distance from $i$ to $j$; and $Y_{IJ}$, the average publication year difference between $i$ and $j$:

$$Y_J^i = \frac{\sum_{j \in T_i : D(j) = J} t_i - t_j}{|j \in T_i : D(j) = J|}.$$

$Y_J^i$ is motivated by previous studies that have examined the number of years taken for drug development [2, 4].

As we are interested in how the three measures vary across source papers with different LS, we randomly selected 101,184 (1%) papers, denoted as $S$, and repeated the above calculations for each $i \in S$, since doing so for millions of papers is computationally burdensome. Aggregating all source nodes, we obtain three matrices, $R_{IJ}$, $L_{IJ}$, and $Y_{IJ}$, respectively representing the average reachability, distance, and publication year difference between nodes with different LS, defined as:

$$R_{IJ} = \frac{\sum_{i \in S: D(i)=I} R_J^i}{|i \in S: D(i) = I|},$$

$$L_{IJ} = \frac{\sum_{i \in S: D(i)=I} L_J^i}{|i \in S: D(i) = I|},$$

$$Y_{IJ} = \frac{\sum_{i \in S: D(i)=I} Y_J^i}{|i \in S: D(i) = I|}.$$

Fig. 5A, which visualizes the matrix $R_{IJ}$, demonstrates the connectivity advantage of basic research. Target papers that located at the basic end of the basic-applied spectrum have a larger portion of paths reaching them regardless of research level of source paper. An applied paper can reach to, on average, 40% of basic papers published before and 20% of applied ones. Fig. 5B shows the matrix $L_{IJ}$. On average, the shortest paths among all sampled pairs of papers happen between those both located at the basic end, with 7.5 steps away from each other. Reaching to papers at the basic end from the applied end requires 10 steps. Fig. 5C plots the matrix $Y_{IJ}$. The publication year difference in an applied-to-basic pair of papers is about 17 years, resonating previous results [4].

# DISCUSSION

The main purpose of this work was to propose a method to place publications onto the translational spectrum, by leveraging recent advances in representation learning. One advantage of our method over previous ones is that rather than grouping papers into different categories, it assigns continuous scores to papers, therefore allowing us to capture the varying "basicness" of papers in the same group. The introduced measure well quantifies research from basic to clinical to health practice. It may also be useful for policymakers to measuring the returns of science investments [28-31].

Throughout the work, we have adopted a working definition that cell and animal related MeSH terms are basic and human related applied. However, not all biomedical research involving human is applied—neuroscience research that advances the understanding of the nervous system using human subject has been considered as basic [32]. Yet, the limitation operates at the term level rather than the paper level, and our bottom-up approach, in that the position of a paper on the basic-applied spectrum is based on the positions of its MeSH terms, may have partially accounted for the limitation. This is based on the consideration that the types of research for papers with the term "Humans" can be diverse, from basic neuroscience research to clinical trial to health practice, and the "basicness" of these papers should not be determined solely by the single term. To support this, Fig. S6 shows papers containing both "Humans" and "Magnetic Resonance Imaging" terms, though still located at the applied side, are more basic than health practice papers.

A related note is about the definitions of basic, translational, and applied research. On one hand, not only the definition of translational science has been evolving [33], but also consensus definitions of basic/applied research have not been reached. Basic research has been defined as, among others, fundamental, curiosity-driven research that leads to scientific discoveries [34], or use-inspired research (e.g., disease-orientated) [32, 35]. On the other, the computation identification of papers from these categories requires some operational definitions, as many previous studies did by looking at journals, title words, or MeSH terms. Here we followed this line and focused on MeSH terms. As such, the proposed level score is simply an attempt to quantify one of the many possible dimensions of translational science. The use of the measure therefore should always be cautious, as any other indicators. When possible, it should be used together with other ways such as a careful reading by domain experts [32]. Future work could also combine level score with other measures, such as whether the research is disease-oriented, and develop multi-dimensional analysis of translational science.

The key idea of our method is the construction of an imaginary Translation Axis that points from basic terms to applied terms. With such a TA, there may be other ways to get the LS of a paper. In Section S2, we provide an alternative one. Specifically, we first get the TA vector as described before and obtain the vector of a paper by averaging the vectors of its MeSH terms. The LS of a paper is then defined as the cosine similarity between the TA vector and the paper vector. Figs. S8-S10 demonstrate that the results remain essentially the same using this formulation.

We relied on MeSH terms assigned to papers. Terms are being updated yearly in various ways, such as additions and deletions. This, however, may not affect our results, since we used a sliding window approach, which examined the usage of terms in papers one period at a time and over different periods.

Limitations remain. First, we examined citation linkages between papers of different LS, looking at the number and length of paths connecting basic and applied papers as well as year difference between them. Although a similar approach has been used to study the development of drugs [6], citations only represent codified knowledge flows and a connected path from basic research to clinical medicine does not mean the occurrence of translation. However, we believe the measures we introduced provide a lower bound. Second, when performing citation analysis, the reliance on WoS for citation data indicates that we missed a significant portion (35.5%) of papers that are indexed in MEDLINE but not presented in WoS. Those papers were excluded, since we have no data about their references. This may to some extent bias our results, but WoS is the most comprehensive data we can access.

Despite these limitations, the availability of a quantity of the degree to which a paper is a basic one paves the way to a number of systematic investigations, opening possibilities for future work. One can, for example, study the association between the level score and the number of citations, understand how funding is allocated along the basic-applied spectrum [36-37], characterize types of research conducted [27] and output of researchers, funding agencies, and research institutes, and examine how papers are cited outside the scientific domain, such as patents, drug products, and clinical guidelines, given the availability of such linkage information. These studies would further advance our understanding of the biomedicine enterprise.

## CONCLUSIONS

We proposed a method to place publications onto the translational spectrum, by learning embeddings of controlled vocabularies. The introduced measure is consistent with previous qualitative results and allows us to characterize "basicness" of journals, fields, and the entire biomedical literature. We found that papers with similar research level tend to cite each other directly, yet papers located at the basic end of the spectrum have the advantage of being more likely to be reached regardless of research level of source paper.

## ACKNOWLEDGMENTS

I thank Pik-Mai Hui for initial collaboration; Cassidy R. Sugimoto for helpful comments to the manuscript; Steven N. Goodman and Griffin M. Weber for discussions; Filippo Radicchi for providing excellent computing resources; Rodrigo Costas for sharing matching data between Web of Science and MEDLINE; the authors of the LINE and GloVe algorithms for open sourcing their code. This work uses the Web of Science data, provided by the Indiana University Network Science Institute.

## COMPETING INTERESTS

None.

## FUNDING

None.

**CONTRIBUTORS**

QK designed research, performed research, analyzed data, and wrote the paper.

**SUPPLEMENTARY MATERIAL**

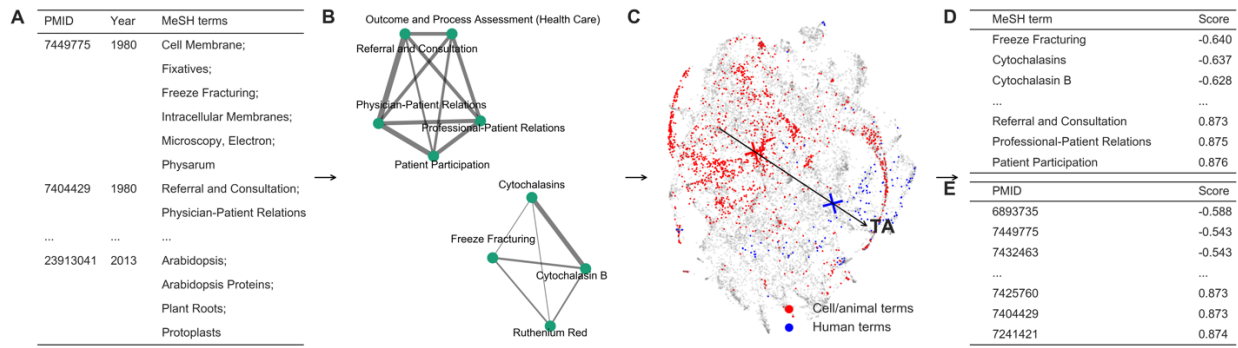

**Figure 1.** Schematic illustration of the calculation of level score (LS) of Medical Subject Headings (MeSH) terms and papers. (A) Papers indexed in MEDLINE are associated with MeSH terms. The table lists selected papers indicated by their PubMed ID (PMID), publication year, and MeSH terms separated by semicolon. (B) For each year $t \in [1980,2013]$, we compute the co-occurrence matrix $M_t$ between MeSH terms based on papers published in the 5-year time window $[t-4, t]$, where the entry $m_{ij}^t \in M_t$ is the number of papers whose MeSH terms contain both term $i$ and $j$. The figure illustrates selected entries in $M_{1980}$ as a network, where we set the edge width as $\ln(m_{ij} + 1)$. (C) We embed each $M_t$ into vector-space, using an embedding method called LINE [17], which assigns each term $i$ a vector $v_i^t$. For illustration, we show their positions at 1980 in 2-$d$ space using t-SNE [19], a dimension reduction technique. In the vector-space, we get the centroid of basic (red dots) and applied terms (blue dots), marked as the large red and blue cross, respectively. Basic terms are those related to cell, molecular, and animal, whereas applied terms related to human [14] (Table S2). We then get the Translational Axis (TA) vector that points from the centroid of basic terms to that of applied terms. The LS of a MeSH term is the cosine similarity between its vector and the TA vector, with larger value indicating the term is more applied. (D) The LS of selected terms in 1980. (E) The LS of a paper is the average of LS of its MeSH terms. The table reports the score of selected

papers. Figs. S11-S18 test the robustness of our results by using another embedding method called GloVe [16], which is widely used for embedding co-occurrence statistics.

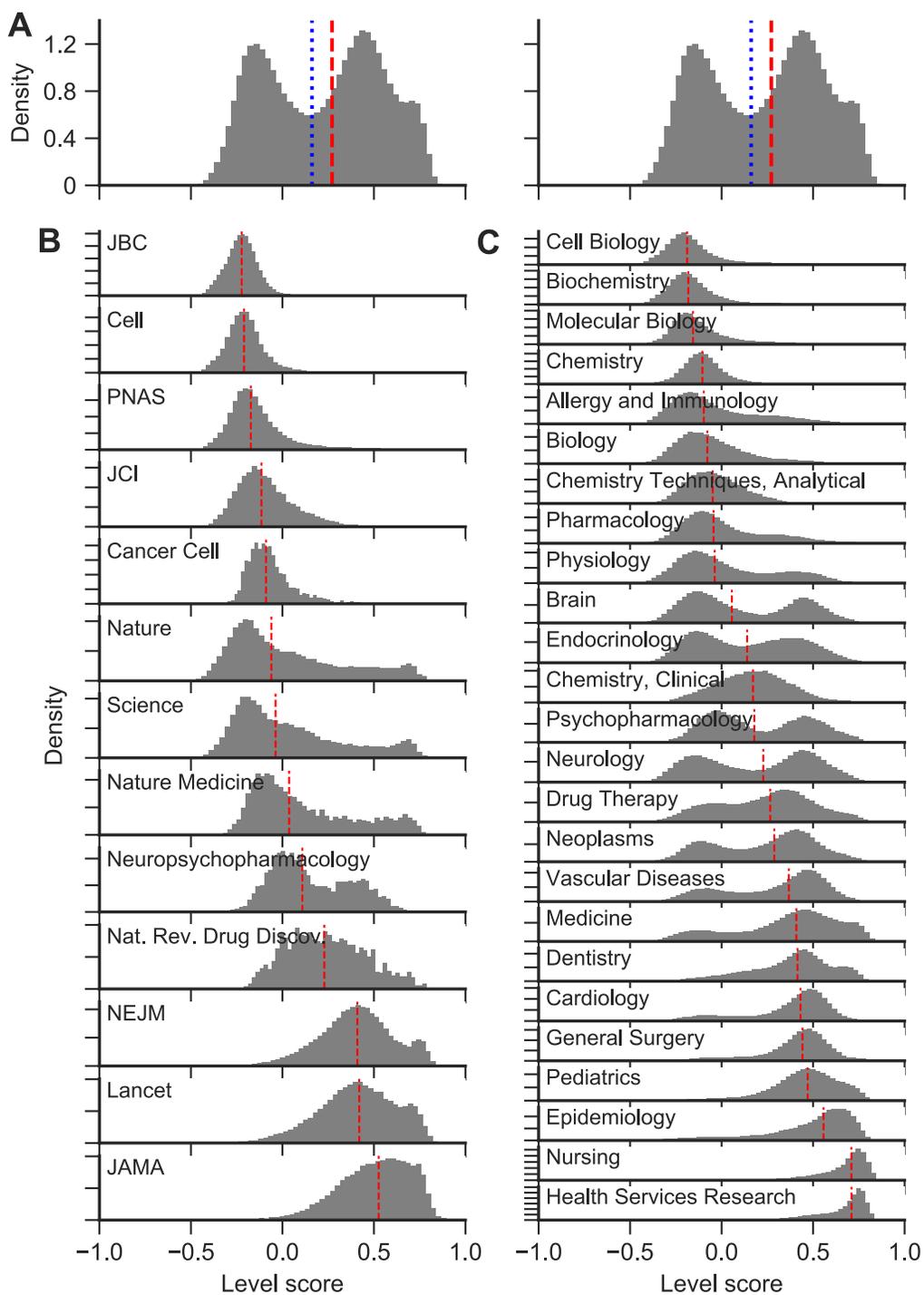

**Figure 2.** Histogram of level score of papers. (A) All MEDLINE papers included in our analysis. The red dashed line marks the median (0.27), and the blue dotted line indicates the score (0.16) corresponding to the local minimum of density, empirically estimated as the midpoint of the bin that achieves such minimum. The right figure is a duplication of the left one, for ease of comparison. (B) Papers published in different journals. JBC: *Journal of Biological Chemistry*; JCI: *Journal of Clinical Investigation*; Nat. Rev. Drug Discov.: *Nature Reviews Drug Discover*; NEJM: *New England Journal of Medicine*. (C) Papers in different disciplines, categorized based on journals where they were published and journal-discipline designations. Each tick on the Y axises represents one unit, with tick labels omitted for clarity.

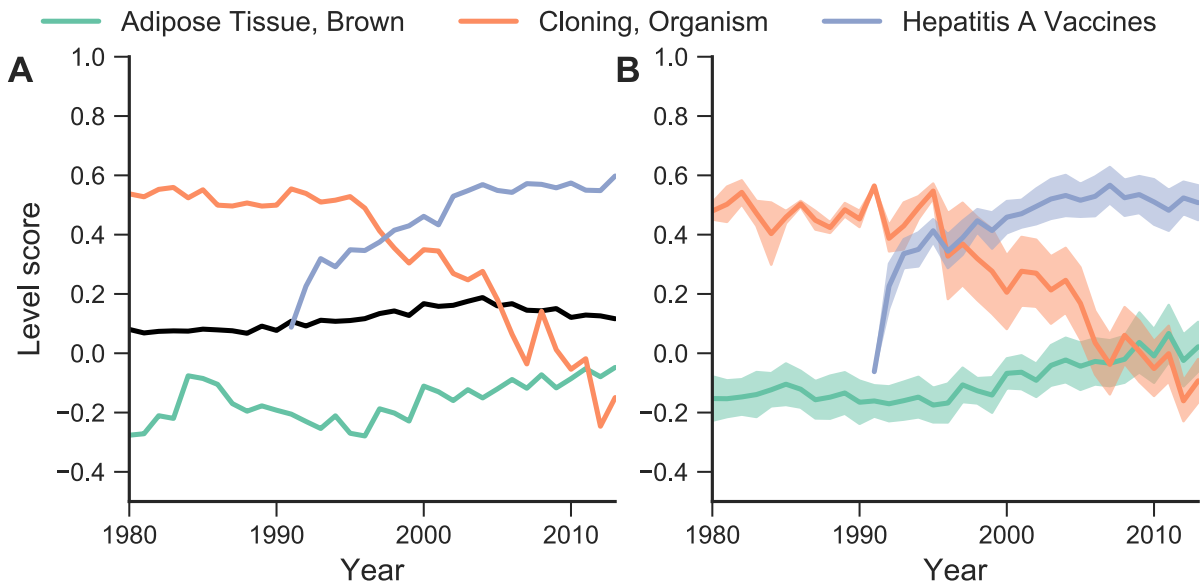

**Figure 3.** Evolution of research level of MeSH terms and papers. (A) LS of different terms. The black line is the average LS of all terms. (B) Mean LS of papers containing the terms. Shaded region covers one standard deviation.

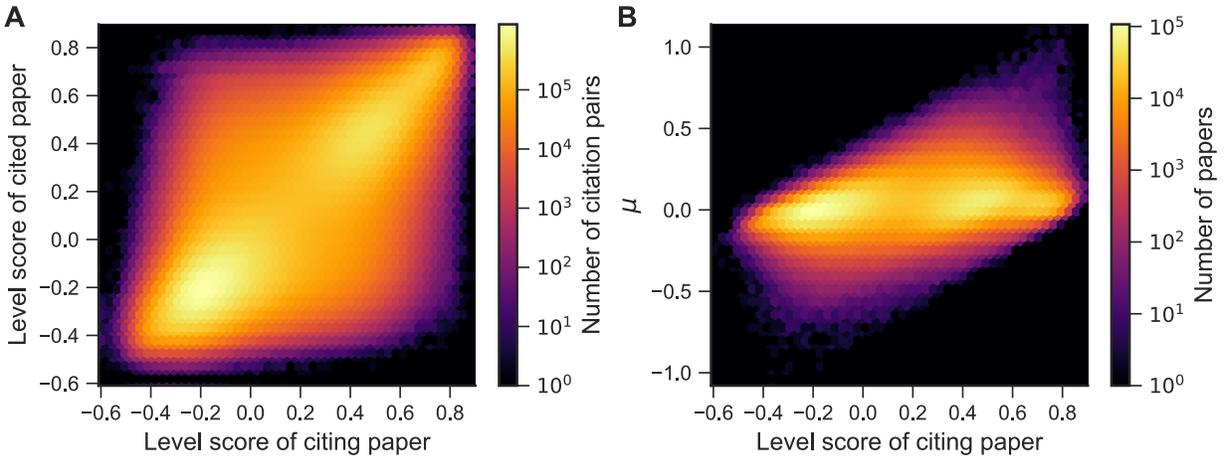

**Figure 4.** Homophilous pairing of citing and cited papers with respect to level score. (A) Heat map of level scores of citing and cited papers, where color encodes the number of pairs. (B) For each paper, we compute the average difference, denoted as $\mu$, between its level score and the scores of its references. The figure shows how $\mu$ is distributed for papers with different scores, where color encodes the number of papers.

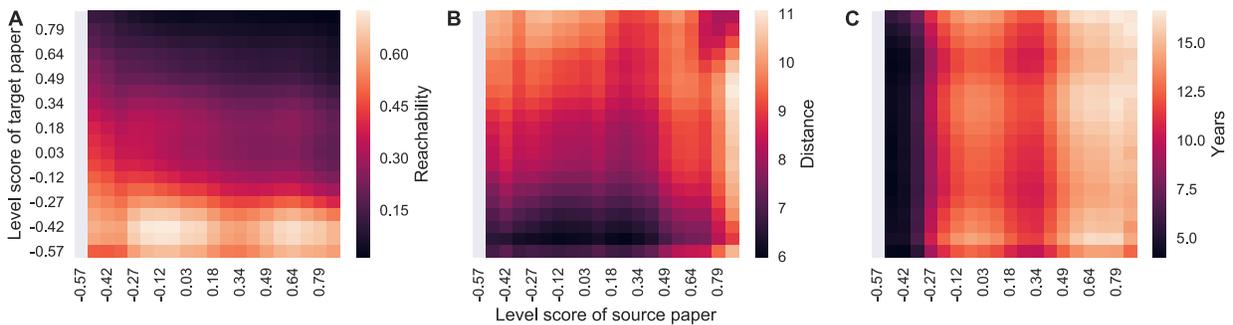

**Figure 5.** Characterization of the citation network with respective to level score. (A) The fraction of source-target pairs of papers where target paper can be reached from the source one. (B) The average path length. (C) The average year differences.